\documentclass{elsart}

\usepackage{graphicx,natbib,amsmath,amssymb}
\usepackage{txfonts}

\def\Msun{M$_{\odot}$}
\def\gtrsim{\mathrel{\hbox{\rlap{\hbox{\lower4pt\hbox{$\sim$}}}\hbox{$>$}}}}
\def\lesssim{\mathrel{\hbox{\rlap{\hbox{\lower4pt\hbox{$\sim$}}}\hbox{$<$}}}}
\def\farcs{\hbox{$.\!\!^{\prime\prime}$}}
\def\ion#1#2{#1$\;${\small\rm{#2}}\relax}
\def\arcsec{\hbox{$^{\prime\prime}$}}

\def\aj{AJ}                   
             
\def\apj{ApJ}                 
\def\apjl{ApJ}                
\def\apjs{ApJS}
\def\aap{A\&A}     
\def\mnras{MNRAS}             
\def\nat{Nature}              
\def\aaps{A\&AS}

\begin{document}

\begin{frontmatter}

\title{On the nature of nearby 
GRB/SN host galaxies\thanksref{labeltitle}}
\thanks[labeltitle]
{Based on observations collected at the European Southern Observatory
Paranal, Chile (ESO Programmes 63.H-0649 and 271.D-5006). 
Also partly based on programmes 70.D-0087 and 165.H-0464. 
Partly based on HST observations obtained under programme 9405.
}

%\author[add1,add2]{J.~Sollerman,}
\author[add1,add2]{J.~Sollerman,\corauthref{cor1}}
\author[add1]{G.~\"Ostlin,}
\author[add2]{J.~P.~U.~Fynbo,}
\author[add2]{J.~Hjorth,}
\author[add3]{A.~Fruchter,}
\author[add2]{K.~Pedersen}

\corauth[cor1]{J.~Sollerman, email: \texttt{jesper@astro.su.se}}

\address[add1]{Department of Astronomy, Stockholm University, 
AlbaNova, 106 91 Stockholm, Sweden}

\address[add2]{DARK, Niels Bohr Institute, University of
Copenhagen, Juliane Maries Vej 30, DK-2100 Copenhagen \O, Denmark}

\address[add3]{Space Telescope Science Institute, 3700 San Martin
Drive, Baltimore, MD 21218, USA }

%\date{Received / Accepted }

\begin{abstract} 
We present and discuss optical diagnostics of the low
redshift ($z<0.2$) galaxies that are known to have hosted supernovae
associated with $\gamma$-ray bursts (GRBs). The three galaxies are all
actively starforming sub-luminous ($L < L^\star$) galaxies with 
relatively low metallicities 
($Z\lesssim Z_\odot$). 
We find no evidence for substantial internal extinction within 
any of the galaxies. 
We derive star formation rates (SFR) based on H$\alpha$ luminosities,
as well as specific star formation rates (SSFR, star formation rate
per unit luminosity). For GRB\,980425 (SN\,1998bw) we use photometry
of the supernova environment to estimate the mass of the progentitor
to $\gtrsim30~M_\odot$.
These three host galaxies have global properties (luminosities, SFR, SSFR, 
metallicity, colour, reddening) that resemble those of more distant GRB host 
galaxies. We also compare the host galaxies with a sample of 
Blue Compact Galaxies (BCGs) in the local universe, and show that these 
samples have similar properties.
\end{abstract}


\end{frontmatter}

\section{Introduction\label{introduction}}

It is now well established that long-duration $\gamma$-ray bursts 
(GRBs) coincide with the explosions of certain massive stars, i.e., with 
a subset of very energetic core-collapse supernovae (SNe).
In particular, three such cases must be regarded as secure. 

The first was the association of 
%\object{SN\,1998bw} with 
%\object{GRB\,980425} \citep{galama98}.
{SN\,1998bw} with 
{GRB\,980425} \citep{galama98}.
This connection was, however, quite debated until 
an unambiguous association was revealed between 
%\object{GRB\,030329} and \object{SN\,2003dh} 
{GRB\,030329} and {SN\,2003dh} 
\citep{hjorth03,stanek03}. SN\,2003dh also showed properties 
almost identical to that of SN~1998bw. A third event has 
now filled in the picture with 
%\object{GRB\,031203} and 
%\object{SN\,2003lw} 
the spectroscopically confirmed association between 
 {GRB\,031203} and 
{SN\,2003lw} 
\citep{thomsen04,malesani04}.

Because the detection of GRBs in $\gamma$-rays is unaffected by intervening 
gas and dust, they provide a powerful and possibly unbiased tracer of 
star-formation in the high-$z$ universe.  This highlights the
importance of studies of GRB host galaxies. Previous studies indicate that 
the majority of GRB host galaxies are blue and sub-luminous 
\citep[e.g.,][]{fruchter99, lefloch03, Jakobsson05}. 

Based on these properties of GRB host galaxies, 
\cite{watson04} discussed whether a sizable portion of
global star-formation occurs in small and rather unobscured, modestly
star-forming galaxies that are too faint to appear in other surveys of
star-formation activity, or whether GRBs trace only a fraction of the
star-forming population, for example due to metallicity effects
\citep[e.g.,][]{fynbo03}.

The most nearby GRB host galaxies, and the ones that are undoubtedly directly
connected to the deaths of massive stars,  allow for a more detailed analysis.
\cite{watson04} used X-ray observations to constrain the star-formation rates
in the host galaxies of the three SN-GRB associations mentioned above.  
In the present
study, we will instead use optical observations to address the
same questions. In Sect.~\ref{observations} we present the data on the hosts
and shortly describe the performed data reductions. In Sect.~\ref{results} we
characterize the host galaxies in terms of emission line fluxes, 
constraints on metallicities and extinctions, star formation rates, and physical dimensions. In
Sect.~\ref{discussion} we discuss and compare the obtained results of
the nearby GRB host 
galaxies with a population of blue compact galaxies. Our findings
are summarized in Sect.~\ref{Conclusions}

A cosmology where $H_0=70$\,km\,s$^{-1}$\,Mpc$^{-1}$, $\Omega_\Lambda = 0.7$
and $\Omega_{\rm m}=0.3$ is assumed throughout.  The redshifts of the three 
host galaxies; SN 1998bw at $z=0.0085$, SN 2003lw at 0.1055 and SN 2003dh at 
0.1685 
then correspond to
%LUMDIST: H0: 70 Omega_m: 0.30 Lambda0 0.70 q0: -0.55 k:  0.00 
%36.6422 487.400      810.327
luminosity distances of 37, 487 and 810 Mpc, respectively.

\section{Observations and data reduction\label{observations}}

\subsection{Spectroscopic data for the GRB\,980425 host\label{spec98bw}}

The host of GRB\,980425 and SN\,1998bw, {ESO\,184-G82}, was extensively
observed as part of several monitoring programmes for
SN\,1998bw.  The evolution of the SN is described in detail elsewhere
\citep{sollerman00,patat01,sollerman02}. 
Here we will make use of some of these
observations to characterize the host galaxy. We have used
the late spectroscopy from 13 June 1999 obtained at the ESO Very Large
Telescope (VLT), %by one of us (JS), 
using the Focal Reducer and low dispersion
Spectrograph (FORS1)
instrument\thanks{http://www.eso.org/instruments/fors1/\\}. 
These data are
presented in \cite{sollerman00} and the spectra 
clearly reveal a number of narrow
emission lines from the underlying host galaxy.  The spectra were obtained
using the 300V and 300I grisms and a $1\farcs0$ wide slit and cover the
wavelength region $\sim$3700--9700~\AA.  
The spectra were reduced in a standard
way, including bias subtraction, flat fielding, and wavelength calibration
using spectra of a Helium-Argon lamp. Flux calibration was done relative to
the spectrophotometric standard star LTT\,7379 \citep{hamuy94}. We did not
only 
use the reduction of \cite{sollerman00}, which aimed at reducing the presence 
of host contamination in the SN spectrum. Instead we re-reduced the data 
to better reveal the underlying emission line region.
We also extracted spectra of two other H\,II regions in the part of the host 
covered by the slit (see Fig.~\ref{f:all}).

The absolute flux calibration of the combined spectrum was obtained relative 
to our well-calibrated broad-band SN photometry.
However, since the slit does not cover the
entire galaxy, we can not derive the total emission line flux and hence
not estimate the global star formation rate. This is instead done from narrow 
band imaging (Sect.~\ref{observationsNB}).

\subsection{Narrow band imaging for the GRB\,980425 host\label{observationsNB}}

For ESO\,184-G82 we have also retrieved narrow band imaging data from the ESO
archive. In particular, we wanted to use the H$\alpha$ imaging to estimate the
global star formation rate of the galaxy. 
These observations were performed on 3 August 2000 with the FORS1
instrument on the VLT. Three 5 minute exposures were obtained in a narrow
filter centered on H$\alpha$ at the host galaxy redshift. Another five minute
exposure was obtained in a zero-velocity narrow H$\alpha$ filter, to allow for
continuum subtraction. The data were reduced in a standard way and the
continuum flux-calibration was performed relative to a photometric standard
star  \citep[PG\,1657,][]{landolt92} observed immediately before the galaxy at
a similar airmass. The broad-band zero-points obtained from this standard star
indicates that the night was photometric.

\subsection{Broad band imaging for the GRB\,980425 host\label{observationsBB}}

We have re-analysed the late time $BVRI$ imaging of the host galaxy of
SN\,1998bw, which were originally used as templates for the SN 
template subtraction photometry in \cite{sollerman02}. 
These are very deep images obtained with FORS1 at a time when the SN 
flux was negligible.  We use it here to construct a 
spectral energy distribution (SED) for 
the galaxy. 
The photometry was calibrated against local photometric standard stars 
\citep{sollerman02}. 
The photometry for the entire galaxy is consistent with the photographic 
ESO-Uppsala catalogue \citep{lauberts89}.

We also did photometry of the small H\,II region that contained the 
SN \citep[see e.g.,][]{fynbo00,sollerman02}.
The magnitude was measured in an aperture with a diameter of 
9 pixels (1\farcs8), 
and the local background was subtracted.

\subsection{Spectroscopic data for the GRB\,030329 host\label{observations1}}

Spectroscopy of the host of GRB\,030329 was obtained on 19 June 2003 
with the FORS2 instrument on the VLT.
As outlined by \cite{gorosabel05} these observations were conducted with the
300V grism and order sorting filter GG375, which efficiently covers the
wavelength range from $\sim$3800--8800~\AA. The 1\farcs3 wide slit 
was used and
the seeing during the observations was below 0\farcs6. Given the small size 
of the object (see Fig.~\ref{f:all}), the bulk of the flux from the host 
galaxy should be included in the slit. Also note that we used a 
position angle of 123.6 degrees East of North, which is well aligned with the
orientation of the host galaxy (Fig.~\ref{f:all}).

The spectra were reduced in the standard way.  Flux calibration was done
relative to the spectrophotometric standard star Feige\,67 \citep{oke90}.  
The absolute flux calibration of the combined spectrum was again obtained 
relative to broad-band photometry. A figure
showing this spectrum is shown by \cite{gorosabel05}.

\subsection{High resolution spectra for the GRB\,030329 host\label{observationsuves}}

We retrieved early high-resolution spectra of GRB\,030329 from the ESO
archive, 
mainly with the aim of searching for the absorption lines of Na~I~D and
hence constrain the amount of extinction along the line-of-sight to the
burst. These lines are often used in SN studies for this purpose 
\citep[see e.g.,][]{sollerman05}.
The observations were obtained 16 hours after the burst on 2003
March 30 with the  Ultraviolet and Visual Echelle Spectrograph
%(UVES)\footnote{www.eso.org/instruments/UVES/\\} on the VLT \citep{greiner03}.
(UVES)\thanks{www.eso.org/instruments/UVES/\\}~on the VLT \citep{greiner03}.
We reduced the spectra using the 
UVES-pipeline\thanks{www.eso.org/observing/dfo/quality/~(vers. 2.0)\\}
~as implemented in $\tt MIDAS$. This reduction package
allows for bias subtraction and flat-fielding of the  data using
calibration frames obtained during day time. Very accurate wavelength 
calibration was secured by comparison to ThAr arc lamp spectra.
From the reduced spectrum we also confirm the redshift reported by
\cite{greiner03} \citep[see also][]{stanek03,hjorth03}, $z=0.1685$.
The [\ion{O}{III}] $\lambda~5006.9$ line was measured at $\lambda~5850.69$, 
which yields
z=0.168525, and  H$\alpha$ $\lambda~6562.8$ was measured at $\lambda~7668.82$ 
which gives z=0.168528. 
The measured wavelengths are corrected for the barycentric velocity.

%--------RESULTS--------------
%

\section{Results\label{results}}

\subsection{Emission line fluxes}

The measured emission line fluxes from the host galaxies of GRB\,980425 and
GRB\,030329 are given in Table~\ref{t:linefluxes}. In that table we also
provide line fluxes for the host galaxy of GRB\,031203, from the 
thorough analysis presented by \cite{prochaska04}.

We note that for GRB\,030329 the main difference between the results presented
by \cite{gorosabel05} and those by \cite{hjorth03} is the absence of
significant
[\ion{N}{II}] emission. The detection of this emission line by \cite{hjorth03}
in the spectrum from 1 May 2003 appears to have been spurious.  It is 
not seen in any of the other spectra published by \cite{hjorth03} and 
the lack of [\ion{N}{II}] emission is also noted by \cite{matheson03}.

\subsection{Metallicities}

From the measured emission line fluxes we have made an 
attempt to constrain the
metallicities of the three host galaxies.  We followed the methods outlined by
\cite{lee03} and by \cite{kewley02} to derive metallicity estimates from the
strong emission lines. 
These estimates are based on the R23 technique which is an empirical relation
between the oxygen abundance and the intensity ratio of the strong oxygen 
emission lines ([\ion{O}{II}] 3727~\AA, and [\ion{O}{III}] 4959,5007~\AA)
to H$\beta$.  This relationship is, however, not unique and therefore provides 
two possible solutions ({\em lower} and {\em upper branch}) for the nebular 
oxygen abundance. This degeneracy may
be resolved by including other lines such as [\ion{N}{II}] 6584 ~\AA.

\subsubsection{The GRB\,030329 host}
Using the prescription of \cite{lee03} for the emission line strengths of
oxygen measured for the host of GRB\,030329 we can derive a metallicity of
either $\sim7.9$ (lower branch) or $\sim8.6$ (upper branch). 
The lower 
branch value would suggest a metallicity significantly below solar
which would also be consistent with the luminosity of the galaxy 
\citep[see, e.g., Fig.~6 in][]{lee03} 
\citep[or Fig.~4 in][]{lee04}. 
However, based on the emission lines alone we can
not exclude a metallicity closer to 
solar \citep[12+log(O/H)$_{\odot}$=8.7,][]{allende01}.

Despite the depth of the VLT/FORS
spectrum, [\ion{O}{III}] 4363~\AA~and [\ion{N}{II}] 6584~\AA~were not 
detected.
Using the various diagnostic tools summarized by \cite{kewley02} we 
did not manage to
break the degeneracy of the oxygen abundance solution based on the R23
ratio.
%An upper limit for [NII]6584 was estimated to be [NII]/H$\alpha$ < 0.036. 
In fact, using our upper limit on \ion{[N}{II]} together with the
[\ion{O}{II}], H$\alpha$, and [\ion{O}{III}] fluxes only gives weak 
constraints
for most published diagnostic diagrams \citep{edmunds84, kewley02}, i.e.,
12+log(O/H)$<8.6$. 
Also our [\ion{S}{II}] detection gives only weak constraints when
compared to \ion{[N}{II]} and H$\alpha$.
Hence we could not constrain the
oxygen abundance of the host of GRB\,030329
using the emission line fluxes, 
but it appears to be sub-solar.

\subsubsection{The GRB\,980425 host\label{980425met}}

For the host galaxy of GRB\,980425 we have  performed a similar
analysis based on the emission line fluxes. We caution that the absolute
fluxing in the bluest part of the spectrum, where [\ion{O}{II}] is located, 
may be affected by systematic uncertainties in the flux-calibration.
This, as well as 
uncertainties in the extinction corrections may affect the metallicity
estimates, but does not alter the overall results.

We extracted spectra from three spatial locations within the host
galaxy (Fig.~\ref{f:all}). The metallicities were
then derived using the recipe in \cite{kewley02}. In this
case, we have enough information to support the upper branch of the
R23 diagnostic, and find the metallicities shown in
Table~\ref{t:metal}. Thus, this galaxy does not have a very
low metallicity. There may be a slight variation in the metallicity
across the galaxy, but given the uncertainties we do not regard this
as a significant result.

We note that the emission lines of [\ion{Ne}{III}] are %significantly
stronger at the site of the burst. As argued by \cite{bloom98} this is
indicative of a high degree of ionization, suggesting a substantial
population of young and massive stars. Such an interpretation is also 
supported by
our modeling of the broad-band spectral energy distribution
(Sect.~\ref{sect:types}).

\subsection{Extinction}

The collected dataset allows some estimates of the extinction in
the hosts. The extinctions are also needed for estimating the 
star formation rates below.

For the GRB\,030329 host galaxy there are several estimates of the extinction
available. First, the photometric spectral energy distribution of the host as
fitted by \cite{gorosabel05} favours a starburst galaxy with an intrinsic
$E(B-V)\sim0.2$~mag. The Galactic extinction in this direction is estimated to
be $E(B-V) = 0.025$~mag \citep{schlegel98}.  \cite{matheson03} also argued for
a low extinction. They based their arguments on the Balmer decrement 
and also made an estimate based on the assumed power-law properties of the
afterglow emission. The latter method gave a limit on the extinction towards
the burst of $E(B-V) = 0.04 \pm 0.08$, implying that there is no
evidence for extragalactic dust along the line of sight between
us and GRB\,030329. 

The high-resolution spectroscopy of the afterglow
of GRB\,030329 discussed above (Sect.~\ref{observationsuves}) 
also speaks in favor of a very low amount of dust towards the GRB.  
At the position of the Galactic sodium lines we do detect a weak line.  
The $\lambda~5890$ component has an equivalent width (EW) of $\sim40$~m\AA.
This small EW is fully consistent with the low
amount of Galactic extinction in this direction 
\citep[e.g.,][]{hobbs74,sollerman05}.
At the position of the Na~I~D
lines at the redshift of GRB\,030329 we detect no significant absorption. 
Any such absorption must be at least four times weaker than the Galactic 
component.
Although the degree of ionization could be different in the GRB\,030329 host
and the Galaxy, this at least points to a very small amount of reddening
along the line of sight to GRB\,030329. 

We must remember, however, that the 
line-of-sight towards the GRB need not be representative for the
entire host galaxy. It may even be that dust is destroyed by the burst 
itself \citep[e.g.,][]{waxman00,galama01,fruchter01}.

For the GRB\,980425 host galaxy we have limited information on
the extinction. The Galactic extinction in this direction is estimated
to be $E(B-V) = 0.059$~mag \citep{schlegel98}. According to
\cite{patat01} there was no sign of Na~I~D absorption in
high-resolution spectra obtained at the SN maximum. These
authors used this finding to argue for an extinction of $E(B-V) <
0.065$~mag towards the SN, i.e., a again very small amount of
reddening. Also the Balmer decrement from the spectra indicate a low 
reddening, which is also consistent with our 
SED modeling (Sect.~\ref{980425mod}).

\subsection{Star Formation Rates}

There are several ways to estimate star formation rates (SFR) in galaxies.
One option is simply to use the emission line luminosities 
of H$\alpha$ or [\ion{O}{II}] and follow the prescription by 
\cite{kennicutt98}.

For the GRB\,980425 host galaxy we have used narrow band imaging to estimate 
the flux in H$\alpha$, since this galaxy is much larger than the extent of 
our spectroscopic slit. 

The final continuum subtracted image is shown in Fig.~\ref{f:all}. 
The host galaxy is clearly visible. We also note that no other source was 
detected in the field of view.
From this image we derived an integrated H$\alpha$ flux of
$2.6\times10^{-13}$~erg~s$^{-1}$~cm$^{-2}$. The integrated EW is 60.6~\AA.  
If we correct these for a 10$\%$ contribution from the [\ion{N}{II}] 
emission line within the filter band (as estimated from the spectroscopy), 
and for a Galactic extinction of $E(B-V)=0.059$ mag we
get for the distance of GRB\,980425 a luminosity of L$_{\rm
H\alpha}=4.4\times10^{40}$~erg~s$^{-1}$. The corrected EW is 54.5~\AA.

We thus derive a global SFR of  $\sim0.35~$\Msun~yr$^{-1}$ 
while the bright H\,II region northwest of the explosion site of 
SN\,1998BW has $\sim0.11~$\Msun~yr$^{-1}$.
These estimates have also been corrected for Galactic extinction and 
for a 10$\%$ 
contamination of [\ion{N}{II}] emission inside the narrow band filter. 

For the GRB\,030329 host we obtain the following integrated SFR;

SFR$_{\rm H\alpha}=0.22~$\Msun~yr$^{-1}$

SFR$_{\rm [O~II]}=0.22~$\Msun~yr$^{-1}$

We note that these values are consistent with the values reported by 
\cite{hjorth03}. The H$\alpha$ line sits close to a telluric absorption line
and was not used by \cite{gorosabel05} to estimate the SFR.
%see sfr.pro
These values are for a Galactic extinction only. 
An extinction as suggested by 
\cite{gorosabel05}, $E(B-V)=0.2$ would alter these numbers to

SFR$_{\rm H\alpha}=0.32~$\Msun~yr$^{-1}$

SFR$_{\rm [O~II]}=0.48~$\Msun~yr$^{-1}$

This of course applies only to the part of the galaxy included in the slit - 
although we argue that this was actually the major part.

These estimates are similar to the SFR reported by \cite{matheson03}, 
0.5~\Msun~yr$^{-1}$, although the agreement appears to be somewhat accidental.
We were unable to reproduce their high L$_{\rm H\alpha}$ also from their 
publicly available spectra, but on the other hand they assume a much lower
extinction correction.

The SFR estimated from the optical emission lines must be regarded as 
lower limits, since there may also exists an extinguished star-forming 
population, as in many other starburts.  For example, the host of GRB\,000210
revealed a SFR $\sim2-3~$\Msun~yr$^{-1}$ from the optical emission lines
\citep{gorosabel03}, while a tentative sub-mm detection implies 
a star formation of several hundred solar-masses per year \citep{berger03}.

In the case of the 
%three
SN/GRB hosts we also have upper limits 
on the star formation as obtained from the unbiased X-ray view exploited by
\cite{watson04}. 
For the GRB\,980425 host they find that the X-ray flux within the optical
extent of this galaxy is entirely dominated by two point sources
$\sim1.5$\arcsec\ apart, one of which is coincident with the radio position of
SN\,1998bw and is almost certainly associated with it
\citep{watson04,kouveliotou04}.  

From the X-ray emission they estimated a total SFR of
$2.8\pm1.9$~\Msun~yr$^{-1}$.  Together with our estimate of
0.35~\Msun~yr$^{-1}$ the range is therefore quite well constrained for this
host galaxy.

%For the other two host galaxies, 
%the limits from the X-ray study are less constraining.
For GRB\,030329 we have estimated SFR$\sim0.4$~\Msun~yr$^{-1}$ while 
\cite{watson04} 
estimate a SFR of massive stars 
(M$\gtrsim5 M_\odot$, see \citealt{watson04,grimm03}) 
of less than 
$31\pm13$~\Msun~yr$^{-1}$.
Using a salpeter IMF \citep{salpeter55,persic04,watson04} 
this corresponds to a
total SFR of $<200\pm80$~\Msun~yr$^{-1}$.

\subsection{Luminosities, Physical sizes and Galaxy types.}\label{sect:types}

\subsubsection{The GRB\,980425 host\label{980425mod}}

The host of GRB\,980425, ESO\,184-G82,  is nearby enough to be well 
resolved with
ground based telescopes (Fig.~\ref{f:all}) and appears to be a late type
spiral with a bar (SBc). The beautiful HST image displayed by \cite{fynbo00}
shows the optical appearance of the galaxy to be dominated by a 
large number
of high surface brightness starforming regions, especially in the southern
spiral arm where the GRB/SN occurred.

The total BVRI magnitudes we obtained from our VLT images 
down to an isophote of 25 magnitudes
arcsec$^{-2}$ are reported in Table~\ref{t:esophot}.  
For the adopted distance
this gives an absolute magnitude of M$_{B}=-17.65$.
Adopting M$_{B}^{\star}=-21$ we thus find that this galaxy has
$L=0.05L^{\star}$.
We can use this and the above derived SFR to also get the
specific star formation rate (SSFR).
For GRB\,980425 we thus derive
a SSFR$\sim7~$\Msun~yr$^{-1}$~(L/L$^{\star}$)$^{-1}$.

The major axis diameter of the galaxy is 67 arcseconds at the B=26.5 mag
arcsec$^{-2}$ isophote, corrected for foreground Milky way extinction, 
i.e. this
is the Holmberg diameter. The minor axis is 57 arcseconds.
At a distance of 37~Mpc this corresponds to a physical size of
$12\times10$~kpc.
%% NGC 4449 (similar to LMC) has 11 kpc

Comparing our broadband magnitudes to empirical galaxy colours
\citep[e.g.,][]{coleman80} we see that the host
appears to be a typical, subluminous late type spiral. To quantitatively 
compare the colors of the host galaxy we calculated a set of models using 
the code PEGASE.2
\citep{pegase,pegase2}
which includes both stellar and nebular emission.
This was done using the actual filter profiles and
CCD sensitivity for FORS1, and all models assume a Salpeter IMF in the mass
interval 0.1-120 \Msun. Using a range of different metallicities and star
formation timescales we could then compare these models to our data using
least-square fitting.

The entire galaxy is well fit with a continuous star formation history. 
Both an e-folding time of 3~Gyr and 15~Gyr give good agreements without 
additional extinction. The results are not very sensitive to metallicity.  
Independently, the H$\alpha$ EW is well fit by such a scenario. 
Both these scenarios are also able to reproduce the integrated luminosity 
of the galaxy given that the current SFR has
operated for 5-7 Gyrs. 
The galaxy can thus not be regarded as a starburst galaxy.

\subsubsection{The GRB 980425 progenitor mass}

For the local SN/GRB environment we have conducted a similar exercise.
The photometry is presented in Table~\ref{t:esophot} and the results of
the modeling are quantified and summarized in Table~\ref{t:ostlin}.
We assumed an instantaneous burst and the same IMF parameters as above.
The best fitting age is for each metallicity a well defined minimum,
where a change in age as small as $\pm 1$~Myr typically increases the
RMS deviation with a factor two or more. For all models, the best-fitting
internal reddening is found to be $E(B-V)=0.05$. As is illustrated in
Table~\ref{t:ostlin}, a change in $E(B-V)$ of $\pm 0.05$, leads to an
increased RMS with typically 50 to $100\%$, but does not affect the best
fitting age. Hence, the uncertainties on the derived ages are very small
($\sim$1 Myr) in a statistical sense.

If we take the best fitting age as the lifetime of the supernova
progenitor we can estimate its mass from comparison with stellar
tracks \citep{bressan93,fagotto94,meyet94}.
This gives a ZAMS mass of 30$\pm$5~\Msun, in accordance
with most models for Type Ic SNe in general, and for collapsars
in particular.

%The PEGASE models are based on stellar evolutionary tracks from the
%Padova group \citep{bressan93,fagotto94}, and therefore our result
%is internally consistent and ultimately depends on the accuracy of
%the stellar tracks.

\subsubsection{The GRB\,030329 host}

For GRB\,030329 the magnitudes reported by \cite{gorosabel05} imply 
$L=0.016L^{\star}$. 
With our measured (extinction corrected) SFR this
gives a SSFR of $\sim25~$\Msun~yr$^{-1}$~(L/L$^{\star}$)$^{-1}$.  
As noted by
\cite{gorosabel05} this is a very high SSFR compared 
to most galaxies in the Hubble Deep Field.

%\cite{gorosabel05} also 
%report that the host appears as a point source on their images. 
To constrain the physical size of this host galaxy we used images
obtained with the ACS onboard the 
Hubble Space Telescope\thanks{programme 9405; P.I. A.~Fruchter.\\}.  
We measured the extent of the host on ACS images obtained on
2004 May 24, i.e., 422 days after the GRB and thus long after the afterglow
contribution had vanished. 

The FWHM of the host is about $530\times930$~pc in the F435W filter image.  
The F606W and F814W filters show a slightly larger FWHM of 
$\sim620\times1030$~pc.

This is similar to the radius estimated by fitting
a Sersic model to
the surface brightness profile \citep[e.g.,][]{warren01}, which gave  
a radius of 0.26 arcseconds corresponding to 750 pc for the F606W filter. We
also estimated the Holmberg diameter to compare directly with the estimate of
the GRB\,980425 host galaxy. 
This gave instead $1\farcs4$,
corresponding to 3.9~kpc.

\begin{figure}[t]
\setlength{\unitlength}{1mm}
\begin{picture}(120,120)(0,0)
\put (0,58)   {\includegraphics[width=57mm,bb=70 226 495 651,clip]{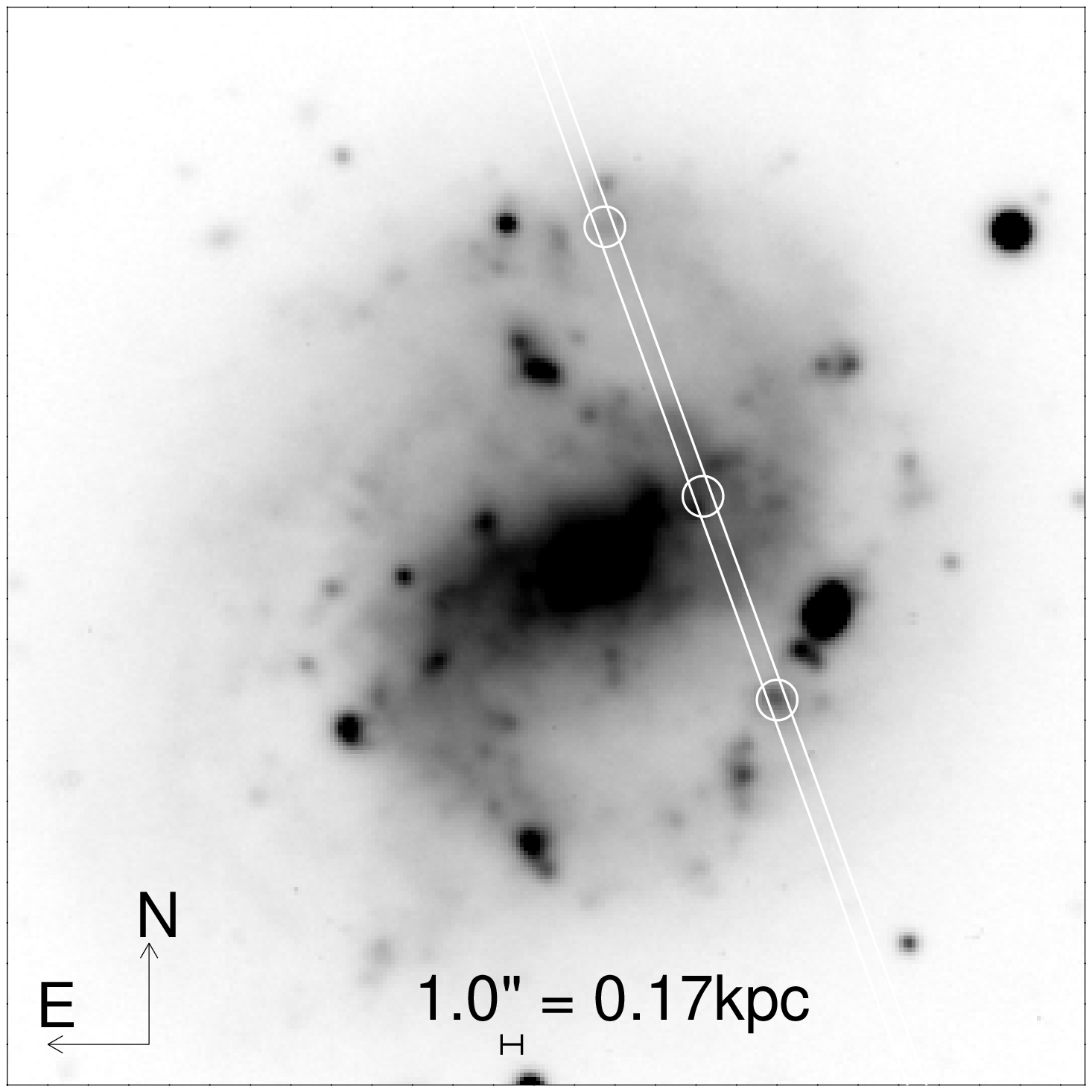}}
\put (58,58)  {\includegraphics[width=57mm,bb=70 226 495 651,clip]{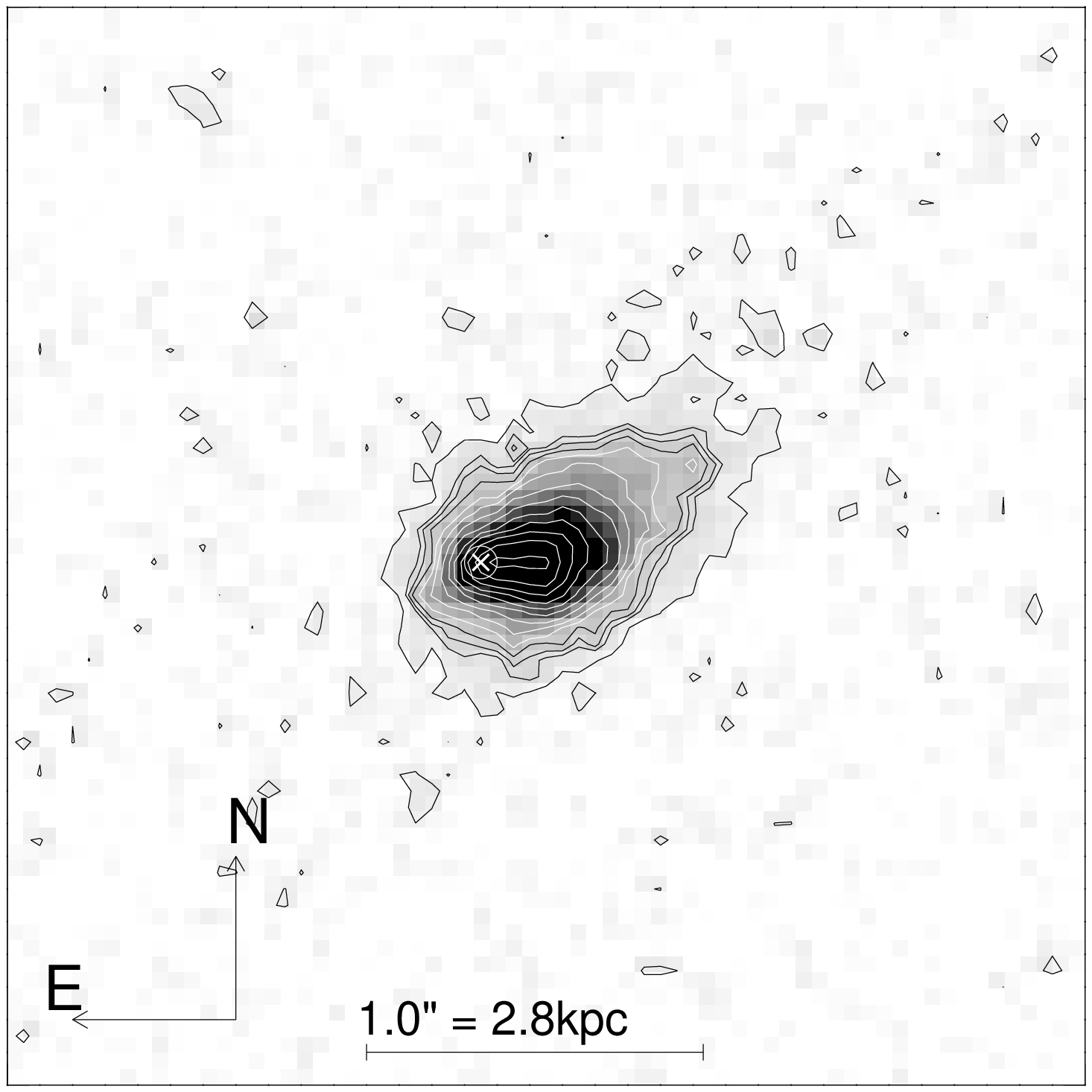}}
\put (58,0)   {\includegraphics[width=57mm,bb=70 226 495 651,clip]{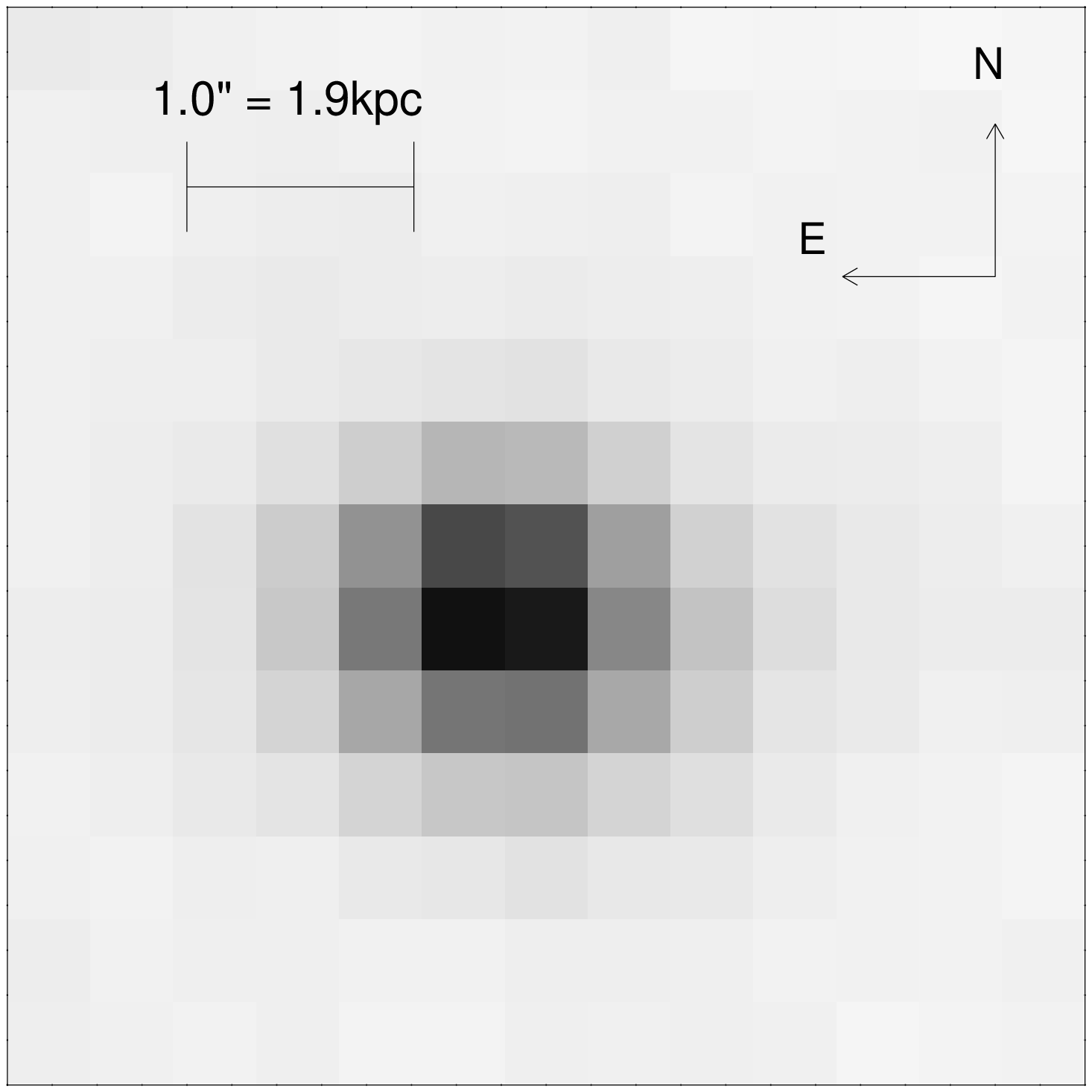}}
\put (0,0) {\includegraphics[width=57mm,bb=70 226 495 651,clip]{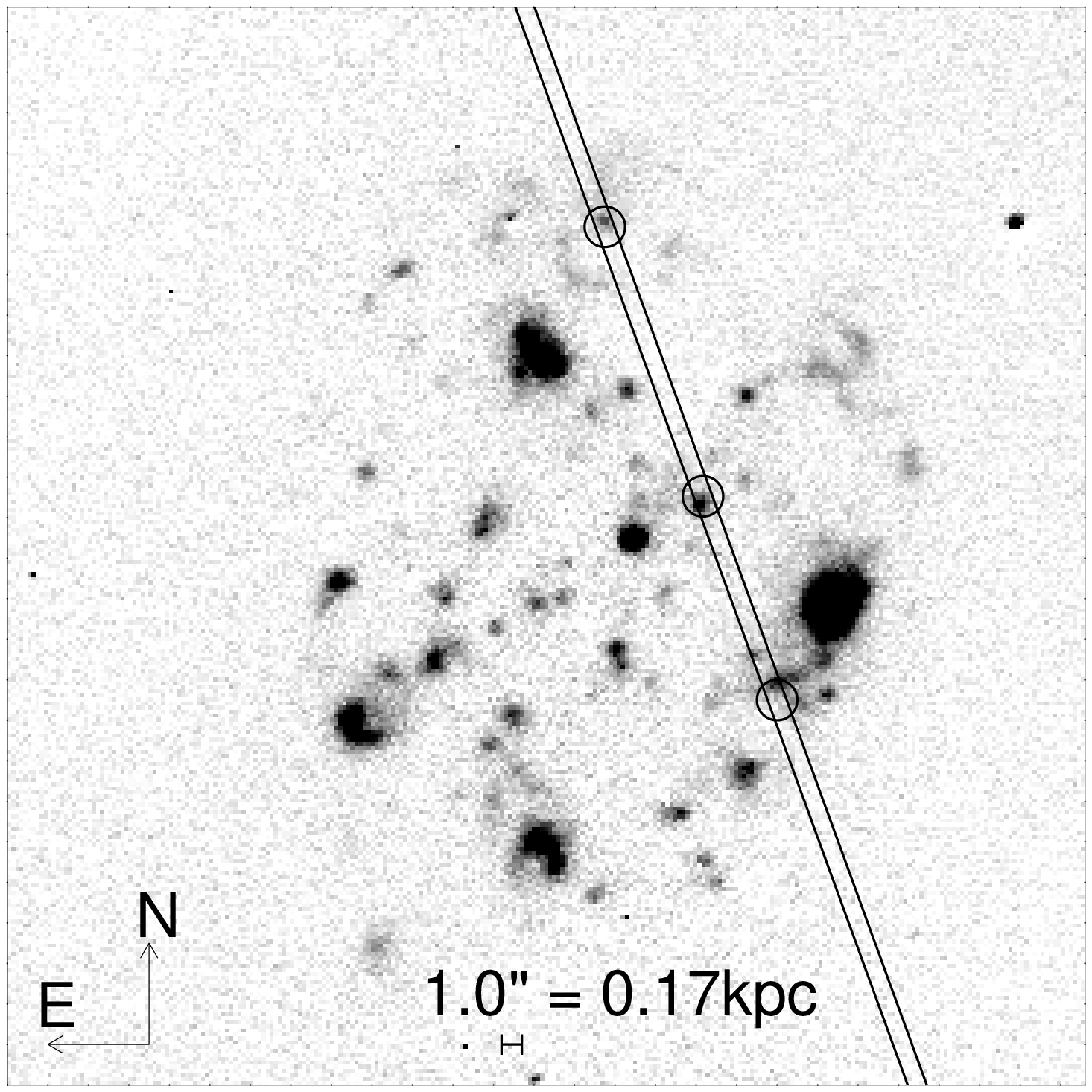}}
\end{picture}
\caption{{\it Upper left:}
The host galaxy of GRB 980425. This is a V-band image obtained with the VLT.
The FOV of this image is $53\farcs2\times53\farcs2$ which corresponds to 9.3~kpc at the distance of this galaxy. The slit position and orientation is marked by white lines. The three regions investigated are marked. The GRB explosion site is the South-Western of these regions.
{\it Upper right:}
The host galaxy of GRB\,030329. This is a F606W image obtained with the HST.
The FOV is the same physical 9.3~kpc as for the GRB\,980425 host shown to 
the left. 
Contours are overlaid the emphasize also the fainter parts of the galaxy.
The position of the GRB is marked by a white ring and a cross.
{\it Lower left:}
This is the H$\alpha$ image of the GRB\,980425 host galaxy. Compare to 
the image above.
{\it Lower right:}
This is an I-band image obtained of the host galaxy for GRB 031203 
with the Danish 1.54m telescope.
The FOV is the same physical 9.3~kpc as for the other host galaxies, but in this case
the galaxy is only marginally resolved.}
\label{f:all}
\end{figure}

The main conclusion of this exercise is that the host galaxy of GRB\,030329 
is a very compact galaxy, with both an absolute magnitude and 
extension similar to that of the SMC.  
Furthermore, our UVES spectroscopy (Sect.~\ref{observationsuves}) 
resolved the H$\alpha$ line with a FWHM of 
$\sim55$~km~s$^{-1}$, which for a radius of 0.75 kpc 
corresponds to a dynamical mass of 
$\sim5\times10^{8}$~\Msun~\citep[e.g.,][]{ostlin01}. 
This may be an underestimate if the H$\alpha$ flux is dominated by 
a central burst and do not trace the full potential well.
The dynamical mass is similar to the mass of SMC.
In Fig.~\ref{f:all} we show the host galaxies of GRBs\,030329 and 980425 
on the same physical scale.
This comparison reveals that these host galaxies are not that different.
The VLT images of GRB\,980425 are considerably deeper and 
surface brightness dimming will also suppress the fainter structures in the
more remote galaxy. One difference is the location of the
GRB within the galaxy. GRB\,030329 appears to have exploded right on the 
brightest pixel in the host galaxy. The location is marked by 
a ring and a cross in
Fig.~\ref{f:all}.  The occurrence of GRB\,980425 is instead way outside the
center of the galaxy, at a projected distance of about 2.2~kpc.
This region does not appear to be special in any way, compared for example to
the very actively star forming region 850~pc northwest of the 
explosion site (Fig.~\ref{f:all}).
We note that 
at cosmological distances such a spatial difference would not have been 
possible to resolve.

From \cite{gorosabel05} we know that the host of GRB\,030329 has a SED that is
best fit by a starburst galaxy template.  We have also measured the 
photometry on the
HST images 
and found m(F435W)=23.29, m(F606W)=23.00 and m(F814W)=22.83.
These are
aperture corrected AB magnitudes. The errors are estimated to be about 0.03 magnitudes. The B and V band results are fairly
consistent with \cite{gorosabel05} although the I band is measured to be
fainter in the HST images.
Looking also at the spatially resolved photometry, we note that the colors
are significantly redder in the outer parts of the galaxy. This means that
using an integrated magnitude will overestimate the
age of the stellar population \citep[e.g.,~][]{ostlin01}. This would make the ages estimated by \cite{gorosabel05} more consistent with the expected ages from very massive GRB progenitors.

\section{Comparisons\label{discussion}}

Having established and collected the properties of the host galaxies of 
GRB\,980425 and GRB\,030329 we will in this section compare these properties 
with those of other host galaxies. First, we compile the properties of the 
third nearby GRB/SN host galaxy, that of GRB\,031203, in the same way as we 
have done for the other two host galaxies. Thereafter we compare this small 
sample of galaxies with a sample of more distant Blue Compact Galaxies.

\subsection{Comparison to the GRB\,031203 host}

The third nearby SN/GRB host galaxy, the host of GRB\,031203, has been
spectroscopically studied in detail by \cite{prochaska04}. 

Using their published reddening corrected emission line fluxes we
performed the same kind of emission line analysis as 
for the other hosts. The resulting metallicities  are 
given in Table~\ref{t:metal}. We note that
our values are somewhat higher than those found by
\citet[12+log(O/H)=8.0]{prochaska04}.
This is probably because the method we have used does not take into account the
additional temperature information from the
fainter [\ion{O}{III}]~4363~\AA~line. 
This makes our abundance 
determinations less secure, and is included here to indicate 
the uncertainties in
the metallicities also for the other
two host galaxies.

When it comes to extinction, \cite{prochaska04} estimated this
from the Balmer decrement to be $E(B-V)=1.17$. Most of
this, $\sim$1.04 mag, is due to extinction in our own galaxy
\citep{schlegel98}. Hence the internal reddening also in this host galaxy is
likely to be modest.

Given this extinction, 
we take the line fluxes from \cite{prochaska04} and use 
the same method as above to infer 

SFR$_{\rm H\alpha}=13.2~$\Msun~yr$^{-1}$, 

SFR$_{\rm [O~II]}=9.6~$\Msun~yr$^{-1}$.

The X-ray data \citep{watson04} suggest a
massive SFR of at most $24\pm17$~\Msun~yr$^{-1}$, 
corresponding to a total SFR
of $\lesssim150\pm110$~\Msun~yr$^{-1}$.  

The low K-band luminosity (L$\sim$L$^{\star}_{K}$/5) 
and the  extrapolated B-band
magnitude \citep{prochaska04} corresponds to 
$L=0.28L^{\star}$. This gives a
SSFR=$\sim39$~\Msun~yr$^{-1}$~(L/L$^{\star}$)$^{-1}$.

For this host galaxy we have limited information about the size
and galaxy type. Using an $I$-band image obtained at the Danish 1.54 m
telescope 46 days past explosion \citep{thomsen04} we compared Gaussian 
fits to the host galaxy with fits of the nearby stars. The galaxy is clearly 
not a
point source, but with a seeing of $\sim1\farcs0$ it is only marginally
resolved. Taking into account the extent of the point spread function this
means that the galaxy FWHM is about 1 -- 2~kpc on the sky. 
This is slightly larger
than the very compact GRB\,030329 host, but not by much (Fig.~\ref{f:all}). 
A J-band image presented by \cite{galyam04} shows that the GRB occurred in 
the central regions of this galaxy.
%Clearly, HST images of
%this galaxy would be valuable to resolve the galaxy and characterize its 
%properties. 

\subsection{Comparison to other galaxies}

The properties of the three SN-GRB host galaxies are summarized in 
Table~\ref{t:big}. The star formation rates given in the table are averages between those obtained for H$\alpha$ and [\ion{O}{II}], when available.
With only three firmly established GRB/SN host galaxies any
conclusion regarding this population will be tentative, but 
it seems
that their overall properties are consistent with those established for more 
distant GRB host galaxies; small, sub-luminous, metal-poor blue galaxies 
\citep[e.g.,][]{fruchter99,lefloch03,fynbo03,christensen04a}. The host 
galaxy of GRB\,980425 has, however, not a very low metal abundance - and 
should not be classified as a starburst galaxy. 
%This differs somewhat from the general conclusions by \cite{christensen04a}.

\cite{christensen04a} 
compared the specific star formation 
rates (UV to optical colours, since the SFRs
were estimated from the near-UV continuum) of GRB hosts
to galaxies in the Hubble deep field, and concluded that the GRB hosts
have higher SSFR than $95\%$ of the HDF galaxies. Such a comparison 
may give the appearance that the GRB hosts have very special properties.

However, the HDF comprises a plethora of galaxies of different types
with different distances and selection functions.
Given that the
hosts under study in the present paper are rather nearby, it would
be interesting to compare these with galaxies in the local
universe with similar properties. 

Although not comprising a very homogeneous class, Blue Compact (dwarf) 
Galaxies (BCGs) and low luminosity emission line  
selected galaxies (H\,II-galaxies) typically have luminosities below 
L$^\star$, low metallicity \citep{izotov99}, 
low extinction  \citep{itl97} and active star formation \citep{ko2000}.
These are properties often ascribed to GRB hosts and the BGC population 
may therefore serve as an interesting local comparison sample.

\begin{figure}[]
\includegraphics[angle=0,width=82mm,clip]{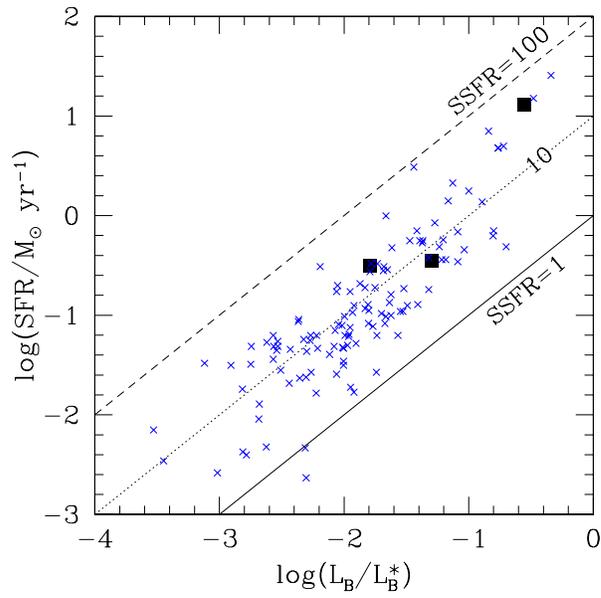}
\caption{Star formation rate vs luminosity. 
The y-axis shows the 
logarithm of the star formation rate 
%in units of $M_\odot$/yr,  
as derived from H$\alpha$, while
the x-axis shows the logarithm
of the blue luminosity.
% in units of $L^\star$
The three GRB hosts marked by solid squares are seen to occupy the same region as the
sample of local blue compact galaxies indicated by cross-symbols. Lines of constant
specific star formation rate (SSFR) are indicated.
}
\label{f:bcg}
\end{figure}

\cite{depaz03} present
visual and H$\alpha$ luminosities for a large  sample of local BCGs, which
we have converted into specific star formation rates. 
We also added a few luminous BCGs with very active star formation
from \cite{ostlin01} to this sample, with the SFRs
recalibrated by the \cite{kennicutt98} SFR--H$\alpha$ relation for the 
sake of homogeneity. In Fig.~\ref{f:bcg} we compare the star formation
rates and luminosities of the combined BCG sample with that of the three 
SN-GRB
hosts of this paper. For this comparison we used the SFRs inferred from
H$\alpha$ only, and since \cite{depaz03} do not present internal extinction,
we use vales corrected for Galactic extinction only in order to be consistent.

The combined samples cover a range in SSFR from $\sim1$ to 100 
\Msun~yr$^{-1}$ (L/L$^{\star}$)$^{-1}$ while the typical BCG has 
a SSFR close to 10. 
While this large range of values
confirms that this class of galaxies is not very homogeneous, 
we see that the SN/GRB host galaxies have similar properties to this class.
It is among these galaxies that we find the most active star
forming sub-luminous galaxies in the local universe.
%(median value 8.4).
The BCGs with SSFR $< 10$ have moderately active star formation and 
this would also be the location occupied by ordinary dwarf irregulars 
and disk galaxies, such as the host of GRB\,980425.
The other two SN/GRB hosts would probably have been classified as actively
starforming BCGs if
they would have been more nearby.

BCGs are believed to account for a significant amount of the SFR also at higher redshifts.
Still, in a magnitude
limited  sample, such  galaxies are in minority. In this
respect the results by \cite{gorosabel05} and \cite{christensen04a} can 
be understood
-- the majority of the (SN/)GRBs occur in the few percent of the
magnitude limited population that has the highest SSFR, both at low
and high redshifts, i.e., galaxies with properties similar to
BCGs/H\,II-galaxies \citep[see also][]{courty04}.

\section{Conclusions\label{Conclusions}}

We have mined several archives and data sets in order to characterize 
the host galaxies that are known to have harbored a GRB-SN. 
The picture of compact low luminosity, metal poorish galaxies in a 
starforming phase is established, consistent with other studies of 
ordinary and more distant GRBs \citep{lefloch03,christensen04a}.
The local extinction in these galaxies appears to be small.
We have proposed that a population of Blue Compact Galaxies have similar 
properties to the GRB-SN host galaxies.

For the host of GRB\,980425 we have for the first time derived the star-formation 
rate. We have shown that this galaxy is not very metal-poor, and that a population
study based on the broad-band photometry is consistent with a normal starforming galaxy with continuous star formation over 5-7 Gyrs, i.e., not with a starburst galaxy. A similar investigation of the spatially resolved H\,II region where the GRB occurred gives us an estimate of the GRB progenitor mass of $\gtrsim30 M_\odot$. This is
consistent with theoretical scenarios of SN\,1998bw being due to a very massive star \citep[e.g.,][]{iwamoto98}. Today, collapsar models - involving the collapse of massive stars - are the favoured models for long GRBs 
\citep[e.g.,][]{macfadyen99,hjorth03}, and it is therefore very encouraging that this population study confirms such models.
Similar observational contraints on the mass of supernova progenitors are becoming very useful for constraining SN models
\citep[see e.g.,][and references therein]{maund05}.
%To our knowledge, this is the first observational constraint on the mass of a GRB/SN progenitor.

For the host of GRB\,030329 we provide high-resolution HST imaging which reveals 
the morphology and location of the burst. It shows the host to be a very compact galaxy indeed. An early high-resolution spectrum provides a very accurate redshift determination and the lack of sodium lines in this spectra supports a very low extinction towards the GRB. The width of the resolved H$\alpha$ provides an estimate of the dynamical mass of the galaxy. Our low-resolution spectra give useful constraints on the metallicity, but we also show that the analysis is less straight-forward than previously acknowledged.

This study thus provides a smorgasbord of what can be learned from optical data of
relatively nearby host galaxies. We hope that more such galaxies will be 
discovered with the 
%recently launched 
{\tt Swift} satellite. 
To characterize this population is important
to understand selection biases in determining the high-z starformation rates
via GRB selected galaxies. {\tt Swift} has the potential to 
point us to a large number of such distant galaxies.

The datasets used in this investigation were mainly
obtained to study the individual SNe-GRBs and were not really optimized for 
the host studies. This is particularly true for the spectroscopy. Clearly, better observations are possible to obtain when the SNe have faded away.

%For GRB\,030329 a deeper study would nail down the metallicity for 
%this galaxy.
%This should aim at detecting also
%[\ion{O}{III}]~4364~\AA~with spectral resolution 
%high enough to separate out [\ion{N}{II}]~6584~\AA~and the nearby telluric
%absorption from H$\alpha$. 

%For the host of GRB\,980425 our spectra have a
%strong SN signal and these observations should be repeated. 
%Deep spectroscopy 
%should 
%aim to detect also [\ion{O}{III}]~4363~\AA. 
%Our photometry revealed a young stellar population close 
%to the site of the explosion, and it would be exciting to 
%confirm  any spatial metallicity gradient in 
%this galaxy, for example using an integral field unit.
%This is the only GRB host galaxy where such a spatial study can be achieved.

%For the host galaxy of GRB\,031203 a very
%detailed spectral analysis was performed by
%\cite{prochaska04}. This demonstrates the usefulness of such comprehensive 
%observations.
%This is the host galaxy with largest SSFR and the lowest metallicity in 
%our sample.
%It would be of interest to obtain spatially resolved broad band photometry of this host using the HST.\\

{\em Acknowledgements.}
We want to thank Javier Gorosabel for discussions about several 
important aspects of this paper.
We also want to thank Lisa Kewley for providing her metallicity script.
Thanks to Christina Th\"one for comments.
Part of this research was conducted at the 
Dark Cosmology Centre
%Danish Dark Universe Centre
funded by The Danish National Research Foundation. 
Important VLT observations were conducted on ESO Director's Discretionary 
Time and were obtained by Paranal staff. We are grateful for these efforts.

{}

\clearpage

%\begin{table}[b]
%\caption{Redshift of the host galaxy of GRB\,030329 from UVES data}
%\label{t:uveslines}
%\begin{tabular}{lccc}
%\hline
%\hline
%      & Rest & Measured$^{a}$ & Redshift \\
%Line  &  (\AA)        &    (\AA)       &           \\
%\hline
%[\ion{O}{iii}]   & 5006.9 & 5850.69 & 0.168525 \\
%H$\alpha$ & 6562.8 & 7668.82 &  0.168528 \\
%\hline
%\end{tabular} \\
%\begin{tabular}{lll}
%\end{tabular}
%$^{a}$ Corrected for barycentric velocity
%\end{table}

\clearpage

\clearpage

\clearpage

\begin{table}[t]
\caption{Emission line ratios}
\label{t:linefluxes}
\begin{tabular}{lccc}
\hline
\hline
Line (\AA) & 980425$^a$  & 030329$^b$ & 031203$^c$  \\ 
%(UT) &         &         &   (s)    &            & (arcsec) \\
\hline
\ion{O}{II} 3727 & 4.54 & 1.61 & 1.06 \\
\ion{Ne}{III} 3870 & 0.40 & 0.22 & 0.60 \\
\ion{Ne}{III} 3970 & 0.24 & -- & 0.23 \\
H$\delta$ 4103 & 0.22 & 0.15 & 0.25 \\
H$\gamma$ 4342 & 0.44 & 0.36 & 0.49 \\
%[OIII] & & --  & 0.11 \\
H$\beta$ 4863 &1.0 & 1.0 & 1.0\\
\ion{O}{III} 4960 &1.07 & 1.12 & 2.11 \\
\ion{O}{III} 5008 & 2.77 & 3.40 & 6.36 \\
He I 5877 & 0.15 & -- & 0.12 \\
%[OI] & & -- & 0.024 \\
%[SII] & & -- & 0.016 \\
%[NII] & & -- & 0.063 \\
H$\alpha$ 6565 & 2.45  & 2.74  & 2.82 \\
\ion{N}{II} 6585 & 0.38 & -- & 0.15 \\
\ion{S}{II} 6726 & 0.89 & 0.40 & 0.16 \\
%[SII] &0.422126 & (incl. above)& 0.071 \\
%[ArIII] & & -- & 0.068 \\
\hline
\end{tabular} \\
\begin{tabular}{lll}
$^a$ Relative to H$\beta$, Corrected only for Galactic extinction of $E(B-V)=0.059$.  \\
$^b$ Corrected only for Galactic extinction of $E(B-V)=0.025$ mag. Total L(H$\alpha$)=2.7$\times~10^{40}$~erg~s$^{-1}$.\\
% sfr.pro
$^c$ GRB 031203 from \cite{prochaska04}, corrected for $E(B-V)=1.17$.
Total L(H$\alpha$)=1.7$\times~10^{42}$~erg~s$^{-1}$.\\
\end{tabular}
\end{table}

\clearpage

\begin{table}[b]
\caption{Host metallicities}
\label{t:metal}
\begin{tabular}{lccc}
\hline
\hline
      & Combined$^{a}$ & R23 Comb  & Comp Ave \\ 
Host  &  & [log(O/H)+12]            &           \\
\hline
031203$^{b}$ & 8.19  &    8.21  &    8.18\\
030329$^{c}$ & 8.65  &    8.65  &    8.67\\ 
980425 SN & 8.62 &     8.43  &    8.56\\
980425  C & 8.61 &     8.67  &    8.73\\
980425 NE & 8.64 &     8.44  &    8.57 \\
\hline
\end{tabular} \\
\begin{tabular}{lll}
$^{a}$ \cite{kewley02} provide details for the various methods.\\
$^{b}$ \cite{prochaska04} found $7.98\pm0.15$ using more lines.\\
$^{c}$ The lower branch value of $\sim7.9$ can not be excluded.\\
%$^{d}$ 
\end{tabular}
\end{table}

\begin{table}[b]
\caption{Photometry for ESO 184-G82}
\label{t:esophot}
\begin{tabular}{lcccc}
\hline
\hline
      & $B$ & $B-V$ & $V-R$  & $R-I$ \\ 
\hline
Region  &          &          &           \\
\hline
Entire galaxy$^{a}$    &  14.94 & 0.40  & 0.40 &  0.41 \\
Local H\,II region       &  22.56 & 0.21  & 0.02 &  $-0.12$   \\

\hline
\end{tabular} \\
\begin{tabular}{lll}
$^{a}$ Magnitude measured down to 25 mag arcsec$^{-2}$. 
Corrected for Galactic extinction only.
\end{tabular}
\end{table}

\begin{table}[b]
\caption{Modeling for the local H\,II region explosion site in ESO-184}
\label{t:ostlin}
\begin{tabular}{lccccccccccc}
\hline
\hline
% \multicolumn{3}{c}{ $E(B-V)=0.0$} & &&  $E(B-V)=0.05$& && $E(B-V)=0.1$\\
& \multicolumn{3}{c}{$E(B-V)=0.0$} & & \multicolumn{3}{c}{$E(B-V)=0.05$} &
& \multicolumn{3}{c}{$E(B-V)=0.10$}  \\
\cline{2-4}
\cline{6-8}
\cline{10-12}
%---- E(B-V)=0.0 ----   --- E(B-V)=0.05 ----    ---- E(B-V)=0.1  -----
Z &     Age &   RMS &   $\Delta_{\rm EW}$ &&    Age &   RMS &
$\Delta_{\rm EW}$ &&    Age &   RMS &  $\Delta_{\rm EW}$\\
&       (Myr) & (mag) & (\AA ) && (Myr) & (mag)& (\AA ) && (Myr) & (mag) &
(\AA ) \\
\hline
0.020 & 6 &  0.100 &  -83 & &   6 &     0.061 & -83 & & 6 &      0.090 &
-83\\
0.010 & 6 &  0.082 &  -51 & &   6 &     0.040 & -51 & & 6 &      0.088 &
-51\\
0.005 & 7 &  0.078 &  -44 & &   7 &     0.055 & -44 & & 7 &      0.102 &
-44\\
\hline
\end{tabular}
\begin{tabular}{lll}
Age: best fitting age in Myr, RMS: root mean square deviation of broad band least square fit, \\ $\Delta_{\rm EW}$ = EW(H$\alpha$)$_{\rm model}$ 
$-$ EW(H$\alpha$)$_{\rm observed}$ in \AA ngstr\"om.
\end{tabular}
\end{table}

%\begin{table}[b]
%\caption{HST photometry for the host galaxy of GRB 030329}
%\label{t:hstphot}
%\begin{tabular}{lccc}
%\hline
%\hline
%      & F435W & F606W & F814W  \\ 
%\hline
%      & 23.29 &  23.00 & 22.83 \\
%
%\hline
%\end{tabular} \\
%\begin{tabular}{lll}
%AB magnitudes, not corrected for extinction.
%\end{tabular}
%\end{table}

\begin{table}[b]
\caption{Properties of the GRB/SN host galaxies}
\label{t:big}
\begin{tabular}{lccc}
\hline
\hline
GRB   &   980425    &    030329   &   031203  \\
SN    & 1998bw      & 2003dh      & 2003lw  \\
\hline
Redshift &   0.0085    &    0.1685   &   0.1055  \\
L$_{B}$/L$^{\star}_{B}$  &   0.05   &   0.016   &   0.28  \\
SFR (\Msun~yr$^{-1}$) &   0.35   &  0.4   &  11 \\
SFR$_{\rm X}$ (\Msun~yr$^{-1}$) &  $2.8\pm1.9$ & $\lesssim200\pm80$ & $\lesssim150\pm110$ \\
SSFR (\Msun~yr$^{-1}$ (L/L$^{\star}$)$^{-1}$) & 7  & 25  & 39 \\
log[O/H]+12 & 8.6  & 8.6$^{a}$  & 8.2 \\
Size (kpc)  &  $12\times10$  & 3.9 & a few \\
$E(B-V)$ (mag) & 0.059 & 0.025-0.2 & 1.17 \\
Age (Myr) & 4500 & 150 & -- \\
\hline
\end{tabular} \\
\begin{tabular}{lll}
$^{a}$ Assuming upper branch
\end{tabular}
%$^{k}$ K-band
\end{table}

\end{document}